\documentclass[twocolumn,preprintnumbers,amsmath,amssymb]{revtex4}

\usepackage{graphicx}
\usepackage[figuresright]{rotating}

\newcommand{\beq}{\begin{equation}}
\newcommand{\eeq}{\end{equation}}
\newcommand{\bea}{\begin{eqnarray}}
\newcommand{\eea}{\end{eqnarray}}
\newcommand{\bdm}{\begin{displaymath}}
\newcommand{\edm}{\end{displaymath}}

\newcommand{\kt}{\textrm{\textbf{k}}}

\def\as{\alpha_s}

\def\ord{{\cal O}}

\def\smallfrac#1#2{\hbox{${{#1}\over {#2}}$}}

\def \msb{\overline{\textrm{MS}}}

\begin{document}
\preprint{MAN/HEP/2010/9} 
\title{High-energy resummation at the LHC: the case of Drell-Yan processes \footnote{
Talk given at $10^{\rm th}$ DESY Workshop on Elementary Particle Theory: Loops and Legs 2010, W\"orlitz, Germany $25^{\rm th}$-$30^{\rm th}$ April 2010.}}
\author{Simone Marzani}
 \email{simone.marzani@manchester.ac.uk}
 
 \affiliation{School of Physics \& Astronomy, University of Manchester,\\Oxford Road, Manchester, M13 9PL, England, U.K.}

\begin{abstract}
We first review the general framework which enables one to resum high-energy logarithms  in hadronic processes, both in the evolution of parton densities  and in the coefficient functions.
We then present an all-order calculation in perturbative QCD of the inclusive Drell Yan and vector boson production in hadron-hadron collisions, in the limit where centre of mass energy is much bigger than the invariant mass of the final state. Our calculation resums leading non-trivial logarithms in the ratio of these two scales.
 We also study the phenomenological relevance of the high-energy corrections for Drell-Yan processes at the LHC. We find that the resummation corrects the NNLO result  by a few percent, for values of the invariant mass of the lepton pair below $100$~GeV. Corrections to $W$,$Z$ production are expected to be of the same order. 
\vspace{1pc}
\end{abstract}

\maketitle

\section{QCD in the high-energy limit}


Perturbative calculations in QCD with initial-state hadrons are possible because of asymptotic freedom and collinear factorization. In particular the latter enables us to separate, in the presence of a hard scale, pertburbative coefficient functions from universal parton densities.
However, when the hard scale $Q$ is
well below the centre-of-mass energy of the colliding hadrons $\sqrt{S}$, the typical values of momentum fraction $x$ of the
colliding partons may be rather small: $x_1x_2 = Q^2/S\ll 1$.  Whenever $x$ is
small, logarithms of $x$ may spoil the perturbative series. Thus
accurate calculations require the computation of the coefficients
of these logarithms, and if they are large it may be necessary to resum them. High-energy logarithms contaminate both the perturbative coefficient functions and the evolution kernels for the parton densities. 

The resummation of small-$x$ logarithms in the perturbative evolution
of parton distribution functions at the next-to leading logarithmic accuracy (NLL$x$) has been performed by different groups
(see for example Ref.~\cite{heralhc} for a comparative review). Any reliable small-$x$ resummation should address the issue of the perturbative instability of the BFKL kernel; the resummed anomalous dimension should be computed at the NLL$x$ accuracy and must match standard DGLAP at moderate values of  $x$. We shall briefly review the resummation procedure proposed by Altarelli, Ball and Forte (ABF), which, in particular, provides a consistent treatment of the factorization scheme dependence. This is an essential ingredient if one wants to match the resummation to fixed order calculations. For a more detailed description of the ABF resummation see  for instance~\cite{Altarelli:2008aj} and references therein.

The general procedure for resumming inclusive hard cross-sections at
the leading non-trivial order through $k_T$-factorization is known~\cite{CCH-photoprod,ch},
and its implementation when the coupling runs understood~\cite{ball}.
High-energy resummed coefficient functions are now known for an increasing number of processes.  Calculations have been performed for photoproduction processes
\cite{CCH-photoprod,ball}, deep inelastic processes \cite{Altarelli:2008aj,ch},
hadroproduction of heavy quarks \cite{CCH-photoprod,ball,ellishq,camici-hq},
and gluonic Higgs production both in the pointlike limit
\cite{Hautmann:2002tu}, and for finite top mass $m_t$~\cite{Marzani:2008az,Marzani:2008ih,Harlander:2009my}. In particular the latter result has been used, together with the $1/m_t$ expansion, to study finite top mass effects in the NNLO inclusive cross-section~\cite{Harlander:2009my,harlanderoz,pak}. The resummation for Drell-Yan and vector
boson production has been performed in~\cite{marzaniballDY} and, more recently, for direct photon production in~\cite{direct}.  Thanks to this effort it is now possible to study the impact of small-$x$ resummation at the LHC.

\section{Drell-Yan and vector boson production cross-section at the LHC}
Accurate perturbative calculations of benchmark processes such as Drell-Yan production of a lepton pair and the closely related  vector boson production cross-section
are an essential component of the LHC discovery programme. Outstanding precision has been reached in the calculation of radiative corrections in perturbative QCD.
The inclusive cross-sections have been known at NNLO for a long time~\cite{Altarelli:1978id,Matsuura:1990ba,Hamberg:1990np,Blumlein:2005im}. The calculation of the rapidity distribution at the same accuracy has been performed in~\cite{NNLOrap}.
More recently the fully differential cross section has also been computed.
The resummation of threshold logarithms
is known up to N$^3$LL \cite{Catani:1989ne,Catani:2003zt,moch,magnea}. 
Thanks to these very precise calculations, the theoretical uncertainty for $W^{\pm}/Z$ production is estimated to be below $4\%$. The dominant contribution comes from parton distribution functions $(1.5-2.5 \%)$~\cite{cteq6.6,mstw08,nnpdf2.0}, while higher order corrections are estimated by scale variations to be below $1\%$. Other sources of uncertainty, such as the treatment of heavy flavours and small-$x$ effects, are more difficult to assess. An estimation of the latter ones is the target of this study.
It is important to control the theoretical calculation at the same precision as the experimental accuracy, which is claimed to be around $4\%$, see for instance~\cite{schott},
so that these processes can be used to monitor the parton luminosity at the LHC. 

Moreover the LHCb experiment aims to measure the Drell-Yan production of a muon pair at invariant mass as low as $2.5$~GeV~\cite{LHCb}. Such a measurement will probe values of $x$ down to $10^{-6}$ and will reduce the uncertainty of the largely unconstrained parton distribution functions at very small values of $x$. On the theoretical side the above mentioned fixed order calculation of the Drell-Yan rapidity distribution shows that the perturbative series is unstable for invariant mass of the muon pair below $20$~GeV, suggesting important contributions from small-$x$ evolution~\cite{Thorne:2008am}.

\section{High-energy resummation}
\subsection{Resummation of the evolution kernels}
The resummation of collinear and high-energy logarithms can be achieved by considering 
the DGLAP and BFKL evolution equations respectively. The DGLAP equations describe the $Q^2$ dependence of the parton densities
 \begin{equation} \label{dglap}
           \frac{ {\rm d} }{ {\rm d} t}G(N,t)= \gamma(N,\alpha_s(t))G(N,t)\,,
 \end{equation}
where $t=\ln \frac{Q^2}{\mu^2}$ and the $N$ is the Mellin moment with respect to $x$. The high-energy, or small-$x$, limit then corresponds to $N\to 0$. We consider the largest eigenvalue $\gamma$ of the anomalous dimension matrix in the singlet sector, which has been computed to NNLO accuracy \cite{singletNNLO}.
The BFKL equation can be written as an evolution equation for the Mellin moments of the unintegrated gluon distribution 
\begin{equation} \label{bfkl}
 \frac{ {\rm d}}{ {\rm d} \xi}\mathcal{G}(\xi,M)=\chi(M,\hat{\alpha}_s(\partial_M))\mathcal{G}(\xi,M)\,,
\end{equation}
where $\xi =\ln\frac{1}{x}$ and $M$ is the
Mellin moment with respect to $Q^2$; therefore the collinear limit in Mellin space is the $M\to 0$ limit.
 The differential operator $\hat{\alpha}_s(\partial_M)$ represents the running coupling in Mellin space; at one-loop we have
 $$
 \hat{\alpha}_s(\partial_M)=\frac{\alpha_s}{1-\alpha_s\beta_0 \frac{\partial}{\partial M}}\,.
 $$
The BFKL kernel $\chi$ has been computed explicitly at the next-to-leading order accuracy
\cite{bfklNLO}. The fixed--order expansion of the kernel is
known to be slowly convergent: the NLO order term $\chi_1$ is large and has
a qualitatively different shape to the leading order kernel $\chi_0$. The origin of this instability is the behaviour of the kernel in the collinear limit: $\chi_0\sim \frac{1}{M}$, while $\chi_1\sim -\frac{1}{M^3}$. The first step towards a reliable evolution at small-$x$ is the resummation of the collinear singularities of the BFKL kernel. In the ABF approach this is achieved by exploiting duality relations among the evolution kernels:
\bea \label{naiveduality} M &=& \gamma(\chi(M,\as),\as), \nonumber \\
 N&=&\chi(\gamma(N,\as),\as)\,. \eea These relations state that at fixed coupling
 the two evolution kernels are the inverse of one another.  Therefore $\chi$ determines the high-energy $(N=0)$
singularities of the DGLAP anomalous dimension and $\gamma$ the
collinear poles ($M=0$) of the BFKL kernel to all orders. Thus we can construct a double leading expansion of the BFKL kernel to next-to leading order:
\beq
\chi^{DL} = \left(\chi_0+ \chi_s -\chi^{d.c.}_0\right)+\left(\chi_1+ \chi_{ss} -\chi^{d.c}_1\right),
\eeq
where $\chi_s$ and $\chi_{ss}$ are dual to the LO and NLO DGLAP anomalous dimension respectively;  therefore collinear logarithms are resummed to NLL. The terms $\chi_{d.c.}^i$ avoid double counting in the two expansions. The double leading kernel, once symmetrized, is stable in the whole region $0<M<1$ and can be used to compute an anomalous dimension which resums both logarithms of $Q^2$ and $x$. However the resummation aims NLL$x$ accuracy and other effects such as running coupling and scheme dependence must be addressed (see Ref~\cite{Altarelli:2008aj} and references therein).
The formalism to consistently resum small-$x$ logarithms in the collinearly factorized coefficient function is discussed in the next sub-section.

\subsection{Resummation of the coefficient function}
\begin{figure}
\begin{center}
\includegraphics[width=0.4\textwidth]{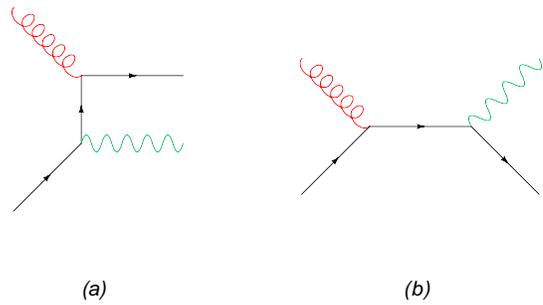}
\caption{Feynman diagrams for the process $ g^*(k) \;  q(p_1) \to  \gamma^* (q) \;q(p_2)$. The quarks are on-shell and massless, while the initial state gluon and the final state photon are off-shell.}\label{Fig:feyn}
\end{center}
\end{figure}
The leading non-trivial high-energy singularities for inclusive cross-sections can be obtained using the $k_T$-factorization theorem~\cite{CCH-photoprod,ch}. 
In \cite{marzaniballDY} we have extended such an approach to deal with the rather complicated flavour structure which characterizes Drell-Yan processes. The main result of that computation is the resummed coefficient function for the quark-gluon sub-process. 
It can be computed by considering the process $$ g^*(k) \;  q(p_1) \to  \gamma^* (q) \;q(p_2) \,,$$ in the framework of  $k_T$-factorization. The relevant Feynman diagrams are shown in Fig.~\ref{Fig:feyn}, where $p_1^2=p_2^2=0$, but $k^2=-|\kt|^2$ and $q^2=Q^2$. The double $N,M$ Mellin transform of the off-shell cross-section, in the small~$N$ limit is found to be
\beq
h_{qg}(0,M) =
\frac{\as}{2 \pi}T_R \frac{4 \,\Gamma(1-M)^2\Gamma(1+M)^2}{(1-M)(2-M)(3-M)}\,.
\eeq
An implicit relation for the resummed coefficient function $D_{qg}$ can be found in terms of the above function $h_{qg}$, the small-$x$ resummed anomalous dimensions $\gamma_{gg}$ and $\gamma_{qg}$ and the scheme dependent factor $R$~\cite{ch}:
\bea\label{Dqgresult}
\gamma_{qg}+ \gamma_{gg}
D_{qg}(N,\as) = h_{qg}(0,\gamma_{gg})
R(\gamma_{gg}) \nonumber \\ +\ord(\as^2(\smallfrac{\as}{N})^k)\,,
\eea

The analytic expression for the leading high-energy behaviour of the coefficient function can be computed to any desired power of the strong coupling. In Mellin space the leading $N \to 0$ behaviour of the $\msb$ coefficient function for the $qg$ channel is
\bea
 D_{qg}(N,\as)= \quad\quad\quad\quad\quad\quad\quad\quad\quad
 \nonumber \\ \frac{\as}{18 \pi} T_R \Big[1
+ \left(\frac{29}{6}
+{2 \pi^2}\right)\frac{C_A}{\pi}\frac{\as}{N}
\nonumber \\
 +\left(\frac{1069}{108}
+\frac{11}{3}\pi^2+{4}\zeta_3 \right)
 \left(\frac{C_A}{\pi}\frac{\as}{N}\right)^2\nonumber \\
 + \left(\frac{9031}{648}+\frac{85}{18}\pi^2+\frac{7}{20} \pi^4
+\frac{73}{3}\zeta_3 \right)   \quad\nonumber \\
\left(\frac{C_A}{\pi}\frac{\as}{N}\right)^3 
 +\dots \Big]. 
\nonumber \\\eea
The coefficients of $\ord \left(\as \right)$ and
$\ord \left(\as^2 \right)$ are in agreement with the
high energy limit of the fixed order NLO \cite{Altarelli:1978id}
and NNLO \cite{Hamberg:1990np,Blumlein:2005im} computations.
This is a very non-trivial check of the procedure. The
$\ord \left(\as^3 \right)$ and subsequent terms are all new results. The high energy singularities of the quark-quark coefficient function
are now easily deduced using the colour-charge relation:
\beq
 D_{qq}(N,\as)= \frac{C_F}{C_A}\left[D_{qg}(N,\as)-\frac{\as}{18 \pi} T_R\right]\,,
\eeq
which is valid in the high energy limit. It can be shown that all the other coefficient functions are sub-leading in the high energy limit~\cite{marzaniballDY}.
At the logarithmic accuracy considered in this study, the high-energy singularities of the vector boson production cross-sections are the same as those for Drell-Yan.
 \section{LHC phenomenology}
\begin{figure*}
\begin{center}
\includegraphics[width=0.3\textwidth, angle=-90]{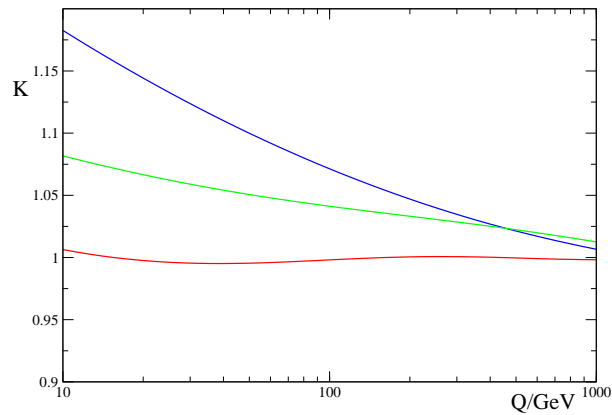}
\caption{In this plot from top to bottom we have: the NLO $K$-factor as defined in Eq.~(\ref{nlokfact}) and the two NNLO $K$-factors of Eq.~(\ref{nnlokfact}), $K^{NNLO}_1$ and $K^{NNLO}_2$ respectively. }\label{Fig:kfact}
\end{center}
\end{figure*}

We now to study the impact of high-energy resummation on the inclusive Drell-Yan cross-section at the LHC. We start by considering the resummed result matched to the fixed order calculation at $\ord(\as)$ using NLO parton distribution functions from the NNPDF collaboration~\cite{nnpdf}, with the evolution performed using the ABF resummed kernel. We define
\beq \label{nlokfact}
K^{NLO}= \frac{D_{ij}^{{\rm NLO }res} \otimes f_i^{{\rm NLO }res} \otimes f_j^{{\rm NLO }res} }{  D_{ij}^{{\rm NLO }} \otimes f_i^{{\rm NLO }} \otimes f_j^{{\rm NLO }}}\,,
\eeq
where $D^{\rm NLO}_{ij}$ are the coefficient functions for the different hard sub-processes and $f_i$'s are the parton distribution functions.
This $K$-factor is plotted in Fig.~\ref{Fig:kfact} as a function of Q (top line in blue) at the LHC centre of mass energy $\sqrt{S}=14$~TeV. At large $Q$ it approaches one, as it should;  at $Q=100$~GeV the resummation corrects the fixed order result by $7$~\% and the effect is as big as $15-20$~\% at $Q =10$~GeV. We then study the effect of the resummation on the NNLO calculation. Because the ABF resummation has not been implemented at this order, we consider two different ratios:
\bea \label{nnlokfact}
K^{NNLO}_1&=& \frac{D_{ij}^{{\rm NNLO }res} \otimes f_i^{{\rm NLO }res} \otimes f_j^{{\rm NLO }res} }{  D_{ij}^{{\rm NNLO }} \otimes f_i^{{\rm NLO }} \otimes f_j^{{\rm NLO }}}\, ,\nonumber \\
K^{NNLO}_2&=& \frac{D_{ij}^{{\rm NNLO }res} \otimes f_i^{{\rm NNLO }} \otimes f_j^{{\rm NNLO }} }{  D_{ij}^{{\rm NNLO }} \otimes f_i^{{\rm NNLO }} \otimes f_j^{{\rm NNLO }}}\,.
\nonumber \\ \eea
This definition of the $K$-factors is such that they again approach one at large $Q$; they are plotted in  Fig.~\ref{Fig:kfact},  $K^{NNLO}_1$  in green in the middle and $K^{NNLO}_2$ in red, bottom curve. 
The effect of the resummation is diminished. In particular $K^{NNLO}_2$ is always very close to one, suggesting the the NNLO coefficient functions capture most of the small-$x$ behaviour of the hard process.  In $K^{NNLO}_1$ instead the high energy contributions to the parton evolution is also taken into account. In this case the resummation corrects the fixed order results by $5-7$\% for $Q<100$~GeV. Thanks to these two results, we can estimate that the resummation corrects the NNLO by no more than a few percent for $Q<100$~GeV. 
Since the effects are rather small, and the choice of parton distribution functions make a sizeable effect on the impact of the resummation, to obtain a reliable calculation of the resummation corrections it will be necessary to refit the PDFs using resummed evolution and cross-sections. 
This is in principle feasible, but it has not yet been done.

The plot also shows that small-$x$ effects are potentially significant for the $W/Z$ cross-section ($Q\sim m_{Z/W}$), where the theoretical uncertainty is claimed to be of the order $3-4\%$. However more detailed studies must be performed: the LL$x$ resummation is the same for Drell-Yan processes and $W/Z$ production but the fixed order results do change.

\begin{figure*}
\begin{center}
\vspace{0.5cm}
\includegraphics[width=0.3\textwidth, angle=-90]{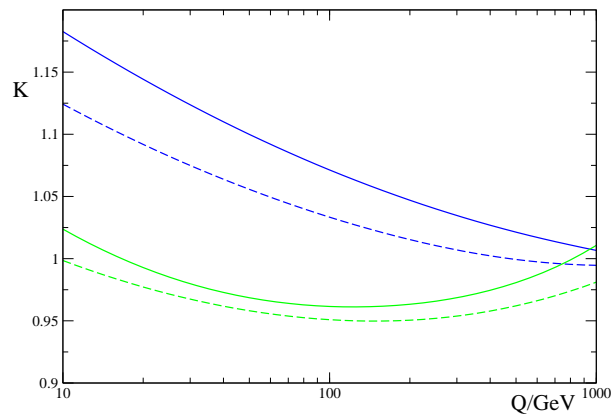}
\caption{In this plot we have the NLO  (top curves) and NNLO (bottom curves) resummed cross-section, normalised to NLO.}\label{Fig:kfactbis}
\end{center}
\end{figure*}

In Fig.~\ref{Fig:kfactbis} we compare the relative size of some of the various contributions beyond NLO. To this purpose we define four $K$-factors with the same denominator: the NLO cross-section computed with NLO partons. The two blue curves at the top are the NLO resummed cross-section with resummed partons (solid line, same as in Fig.~\ref{Fig:kfact}) and the NLO resummed cross-section with standard NLO partons (dashed line). The curves at the bottom are the NNLO resummed cross-section computed with NNLO parton distributions (solid green) and NLO ones (dashed green). 
The gluon-gluon subprocess makes a substantial contribution at NNLO. However this contribution has not been resummed yet: to do this would require the calculation of the off-shell process $g^* g^* \to \gamma^* q\bar{q}$.

\section{Conclusions}
We have presented analytic results for the Drell-Yan (and thus also vector boson production) 
coefficient functions to arbitrarily high orders in the strong coupling,  in the limit of high partonic centre-of-mass energy. Our results are given in $\msb$ scheme and they agree with the known results at NLO and 
NNLO, while providing new results at N$^3$LO and beyond.

We have evaluated the effect of high-energy resummation by computing K-factors between resummed and standard fixed order results. We have considered proton-proton collision at $\sqrt{S}=14$~TeV. We have found that the resummation corrects the NLO result by $5$-$10$ \% for $Q< 100$~GeV. When NNLO corrections are included the effect is reduced to a few percent in the same kinematical region. 
More studies have to be done, especially for vector boson production, where the effects of small-$x$ resummation could potentially be of the same order of the current  theoretical uncertainty.
 \vspace{0.3cm}
 



\begin{footnotesize}

\end{footnotesize}


\end{document}